\documentclass[pra, aps]{revtex4}  

\usepackage[T1,T2A]{fontenc}
\usepackage[english]{babel}
\usepackage{amsmath,amsfonts,amssymb}
\usepackage{graphicx}

\newcommand{\Tr}{\mathrm{Tr}}

\begin{document} 

\title{Quantum entanglement and quantum discord in dimers in multiple quantum NMR experiments}

\author{S. A. Gerasev$^{1, 2}$, E. I. Kuznetsova$^1$}
\affiliation{$^1$ Institute of Problems of Chemical Physics, Chernogolovka, Moscow Region,
142432, Russia}
\affiliation{$^2$ Lomonosov Moscow State University, Faculty of Fundamental Physical-Chemical Engineering,119991,Moscow GSP-1, RUSSIA}

\begin{abstract}
Quantum correlations in multiple quantum (MQ) NMR experiments are investigated in two-spin systems (dimers). In the initial moment of time one spin is in a pure quantum polarized state and the other spin is in the thermodynamic equilibrium state defined by the temperature of the sample. MQ NMR dynamics  of dimers is investigated. It is shown that MQ NMR coherences of only the zeroth and second orders emerge in such a system. The intensities of those coherences are calculated. Entangled states appear in the course of the system evolution in the MQ NMR experiment. The quantum discord is obtained in the high temperature approximation.  
\end{abstract}


\maketitle

\pagestyle{empty} 

\section{Introduction}
\label{sec:intro}  

Multiple  quantum (MQ) NMR \cite{baum} is an effective method for investigation of various problems of quantum informatics. MQ NMR creates multiple-qubit coherent states which can be used for the investigation of the quantum state transfer \cite{cappelaro}, quantum correlations \cite{pyrkov,kuznetsova}, and decoherence processes on a basis of the study of relaxation of MQ NMR coherences \cite{doronin,kaur,bochkin}.
 
Dimers are very suitable systems for a study of quantum correlations. Crystalline hydrates are examples of such systems. It is important that MQ NMR dynamics can be investigated analytically at different initial states of the system \cite{furman}. 

Dimers allow an investigate of not only entanglement, which is a measure of quantum correlations in pure states \cite{zenchuk}, but also the quantum discord \cite{henderson,zurek}  which is a measure of quantum correlations both in pure and mixed states. Those measures of quantum correlations are closely connected with the intensities of MQ NMR coherences which can be observed in  MQ NMR experiments \cite{baum}.

The paper is organized as follows. MQ NMR dynamics of dimers is investigated in section 2. It is shown that MQ NMR coherences of the zeroth and second orders only emerge in dimers. The intensities of those coherences are calculated. In section 3 we show that entangled states appear in the course of the system evolution in the MQ NMR experiment with dimers. We find that the concurrence \cite{hill,wootters} which is the important characteristics of entanglement is connected with the intensity of the MQ NMR coherence of the second order. The quantum discord in dimers is investigated in the high temperature approximation in section 4. The optimization of the quantum conditional entropy is performed with ''Mathematica''. We briefly summarize our main results in section 5.

\section{MQ NMR dynamics of dimers}
We consider a system of two spins connected by the dipole-dipole interaction in an external magnetic field. Initially one spin is in a pure state and the second spin is in a thermodynamic equilibrium state. The initial density matrix $\rho_0$ of the system can be written as
\begin{equation}\label{rho0}
\rho_0=
\dfrac{(\alpha|0\rangle+ \beta|1\rangle)(\alpha^{*} \langle 0|+\beta^{*} \langle 1|) \otimes e^{\frac{\hbar \omega_0}{kT} I_{z,2}}}{Z} 
\end{equation}
where $|0\rangle$, $|1\rangle$ form a basis in the Hilbert space of the first spin, $\alpha$, $\beta$ are complex numbers ($|\alpha|^2+|\beta|^2=1$), $\omega_0$ is the Larmor frequency. $T$ is the temperature, $I_{z,2}$ is the $z$-projection operator of the second spin, and $Z$ is the partition function ($Z=\Tr\{e^{\frac{\hbar \omega_0}{kT} I_{z,2}}\}$). In the standard basis $|00\rangle$, $|01\rangle$, $|10\rangle$, $|11\rangle$ \cite{nielson} the density matrix \eqref{rho0} can be written as
\begin{equation}\label{rho0matrixform}
\rho_0= \begin{pmatrix}  
\frac{e^b|\alpha|^2}{e^b+1}
& 0  & \frac{e^b\alpha\beta^*}{e^b+1} & 0 \\
0  & \frac{|\alpha|^2}{e^b+1} & 0 & \frac{\alpha\beta^*}{e^b+1} \\
\frac{e^b\beta\alpha^*}{e^b+1} & 0 & \frac{e^b|\beta|^2}{e^b+1} & 0 \\
0 & \frac{\beta\alpha^*}{e^b+1} & 0 & \frac{|\beta|^2}{e^b+1}   
\end{pmatrix},
\end{equation} 
where $b=\frac{\hbar \omega_0}{kT}$. For a two-spin system the Hamiltonian $H_{MQ}$, describing MQ NMR dynamics on the preparation period of the MQ NMR experiment \cite{baum}, can be presented in the following form
\begin{equation}\label{mq-hamiltonian}
H_{MQ}=d \left(I_1^+I_2^+ + I_1^-I_2^- \right),
\end{equation}
where $d$ is the dipolar coupling constant and $I_i^+$ and $I_i^-$ are the raising and lowering operators. In the end of the preparation period of the MQ NMR experiment the density matrix $\rho(\tau)$ is 
\begin{equation}\label{resultliuvill}
\rho \left( t\right) =e^{-iH_{MQ}\tau}\rho(0)e^{iH_{MQ}\tau}.
\end{equation}
Using the technique of Ref. \cite{pyrkov} one can obtain the following expression for the density matrix
\begin{equation}\label{rhotaumatrixform}
\rho(\tau)= \begin{pmatrix}  
\frac{e^b|\alpha|^2\cos^2(d\tau)+|\beta|^2\sin^2(d\tau)}{e^b+1}
& -\frac{i\beta\alpha^*\sin(d\tau)}{e^b+1}  & \frac{e^b\alpha\beta^*\cos(d\tau)}{e^b+1} & \frac{i\sin(2d\tau)(e^b|\alpha|^2-|\beta|^2)}{2(e^b+1)} \\
\frac{i\beta^*\alpha\sin(d\tau)}{e^b+1}  & \frac{|\alpha|^2}{e^b+1} & 0 & \frac{\alpha\beta^*\cos(d\tau)}{e^b+1} \\
\frac{e^b\alpha^*\beta\cos(d\tau)}{e^b+1} & 0 & \frac{e^b|\beta|^2}{e^b+1} & \frac{ie^b \beta\alpha^*\sin(d\tau)}{e^b+1} \\
-\frac{i\sin(2d\tau)(e^b|\alpha|^2-|\beta|^2)}{2(e^b+1)} & \frac{\beta\alpha^*\cos(d\tau)}{e^b+1} & -\frac{ie^b \beta^*\alpha\sin(d\tau)}{e^b+1} & \frac{e^b|\alpha|^2\sin^2(d\tau)+|\beta|^2\cos^2(d\tau)}{e^b+1}   
\end{pmatrix}.
\end{equation} 
The diagonal elements $\rho_0(\tau)$ of the density matrix \eqref{rhotaumatrixform} are responsible for the MQ NMR coherence of the zeroth order. The non-diagonal elements $\rho_2(\tau)$, $\rho_{-2}(\tau)$ of $\rho(\tau)$ are responsible for the MQ coherences of the plus/minus second orders. It means that the density matrix $\rho(\tau)$ can be represented as
\begin{equation}\label{components_of_densitymatrix}
\rho(\tau)=\rho_0(\tau)+\rho_2(\tau)+\rho_{-2}(\tau).
\end{equation}
It is natural that in two-spin systems only MQ NMR coherences of the zeroth and second orders emerge. An analogous situation takes place \cite{feldlacele,feldlacele2,maximov,feldman} in linear spin chains in the approximation of the nearest neighbor dipolar interactions \cite{goldman}. The intensities of the MQ NMR coherences of the zeroth ($G_0(\tau)$) and plus/minus second ($G_{\pm 2}(\tau)$) orders, which can be observed experimentally, are determined by
\begin{equation}\label{intensities}
\begin{array}{l}
G_0(\tau)=\Tr\{\rho_0(\tau)\rho_0^{ht}(\tau) \},\\
G_{\pm 2}(\tau)=\Tr\{\rho_2(\tau)\rho_{-2}^{ht}(\tau) \},
\end{array}
\end{equation}
where $\rho_0^{ht}(\tau)$ and $\rho_{\pm 2}^{ht}(\tau)$ are the diagonal and non-diagonal parts of the high temperature density matrix \cite{feldman2002}
\begin{equation}\label{htmatrix}
\rho^{ht}(\tau)=\rho_0^{ht}(\tau)+\rho_2^{ht}(\tau)+\rho_{-2}^{ht}(\tau)=e^{-iH_{MQ}\tau}I_z e^{iH_{MQ}\tau},
\end{equation}
where $I_z=I_{1z}+I_{2z}$. Tedious calculations with \eqref{rhotaumatrixform}-\eqref{htmatrix} allow us to obtain
\begin{eqnarray}
G_0(\tau)=
\frac{e^b|\alpha|^2-|\beta|^2}{e^b+1}\cos^2(2d\tau), \label{intensitie2g0}\\
G_{\pm 2}(\tau)=
\frac{e^b|\alpha|^2-|\beta|^2}{2(e^b+1)}\sin^2(2d\tau).\label{intensitie2g2}
\end{eqnarray}

\section{Entanglement of dimers in MQ NMR experiments}

In order to calculate  entanglement in dimers in MQ NMR experiments we use the criterion of \cite{wootters}. To this end we consider the matrix
\begin{equation}\label{matrixforconcurrence}
\tilde\rho(\tau)=(\sigma_y\otimes\sigma_y)\rho^*(\tau)(\sigma_y\otimes\sigma_y),  
\end{equation}
where $\sigma_y$ is the Pauli matrix. The concurrence, which is closely connected with entanglement \cite{hill,wootters}, is determined \cite{wootters} by the relation
\begin{equation}\label{lambdasforconc}
 C =\max\{0, 2\lambda-\lambda_1-\lambda_2-\lambda_3-\lambda_4 \}, \; \lambda=\max\{\lambda_1,\lambda_2,\lambda_3,\lambda_4\}, 
 \end{equation}
where $\lambda_1,\lambda_2,\lambda_3,\lambda_4$ are the square roots of the eigenvalues of the matrix product $\rho(\tau)\tilde{\rho}(\tau)$.

   \begin{figure} [ht]
   \begin{center}
   \begin{tabular}{c} 
   \includegraphics[height=5cm]{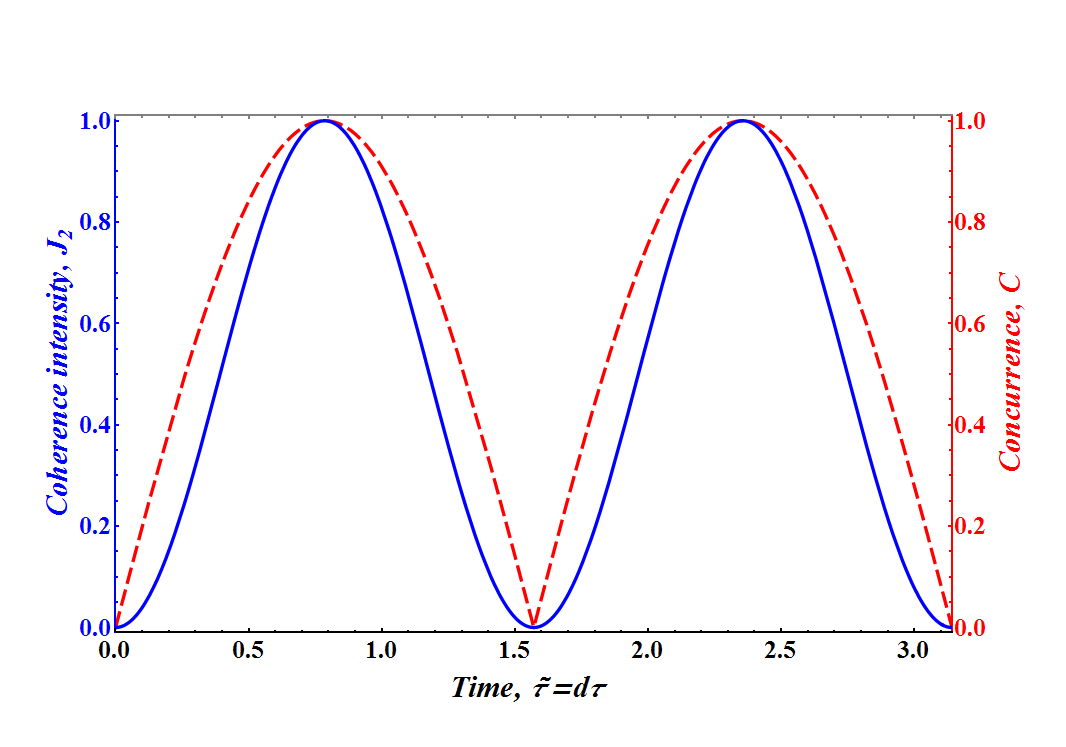}
	\end{tabular}
	\end{center}
   \caption[example] 
   {\label{fig:risunok1} 
The dependencies of the intensity of the MQ NMR coherence of the second order ($J_2(\tau)=G_2(\tau)+G_{-2}(\tau)$) and the concurrence $C$ on the dimensionless preparation period time $\bar{\tau}=d\tau$ at $b=10, \alpha=1, \beta=0$.}
   \end{figure} 

After tedious calculations one obtains
\begin{equation}\label{concurrence}
C=\frac{|(e^b|\alpha|^2-|\beta|^2)\sin(2d\tau)|}{e^b+1}
\end{equation}
Comparing Eq. \eqref{intensitie2g2} and Eq. \eqref{concurrence} one can find an important connection of the concurrence and the intensities of the MQ NMR coherence of the second order
\begin{equation}\label{g2concurrence}
C=\sqrt{\frac{|(e^b|\alpha|^2-|\beta|^2)(G_{2}(\tau)+G_{-2}(\tau))|}{e^b+1}}.
\end{equation}
Such a comparison is shown in Fig.1.

\section{The quantum discord in the high temperature approximation}

According to the  current understanding \cite{aldoshin}, total (quantum and classical) correlations in a system are defined by the mutual information. The problem is how to separate the classical correlations from quantum ones. The problem was solved in \cite{henderson,zurek}. The classical correlations in a bipartite system are determined by a complete set of projective measurements carried out only over one of the subsystems \cite{zurek}. We take the corresponding projectors $\Pi_+$, $\Pi_-$ in the form \cite{doronin2015}
\begin{equation}\label{projectors}
\begin{array}{l}
\Pi_+=
\frac{1}{2}+\frac{n_xI_x+n_yI_y+n_zI_z}{2},\\
\Pi_-=
\frac{1}{2}-\frac{n_xI_x+n_yI_y+n_zI_z}{2},
\end{array}
\end{equation}
where $I_{\alpha}$ ($\alpha=x,y,z$) is the spin projection operator and the parameters (the directional cosines) $n_x, n_y, n_z$ are connected by the relation
\begin{equation}
n_x^2+ n_y^2+ n_z^2=1.
\end{equation}
After performing projective measurements over the density matrix \eqref{rhotaumatrixform} we can calculate the quantum conditional entropy \cite{doronin2015}. However, this entropy depends on the parameters $n_x, n_y, n_z$. At the same time, we need to find the minimal value of the quantum conditional entropy. We solved the problem only in the high temperature approximation \cite{goldman} when $b<<1$. Using ''Mathematica'' we obtain that the entropy reaches it minimal value at 
\begin{equation}\label{usloviyamin}
n_x=0,\; |n_y|=1,\; n_z=0.
\end{equation}
   \begin{figure} [ht]
   \begin{center}
   \begin{tabular}{c} 
   \includegraphics[height=5cm]{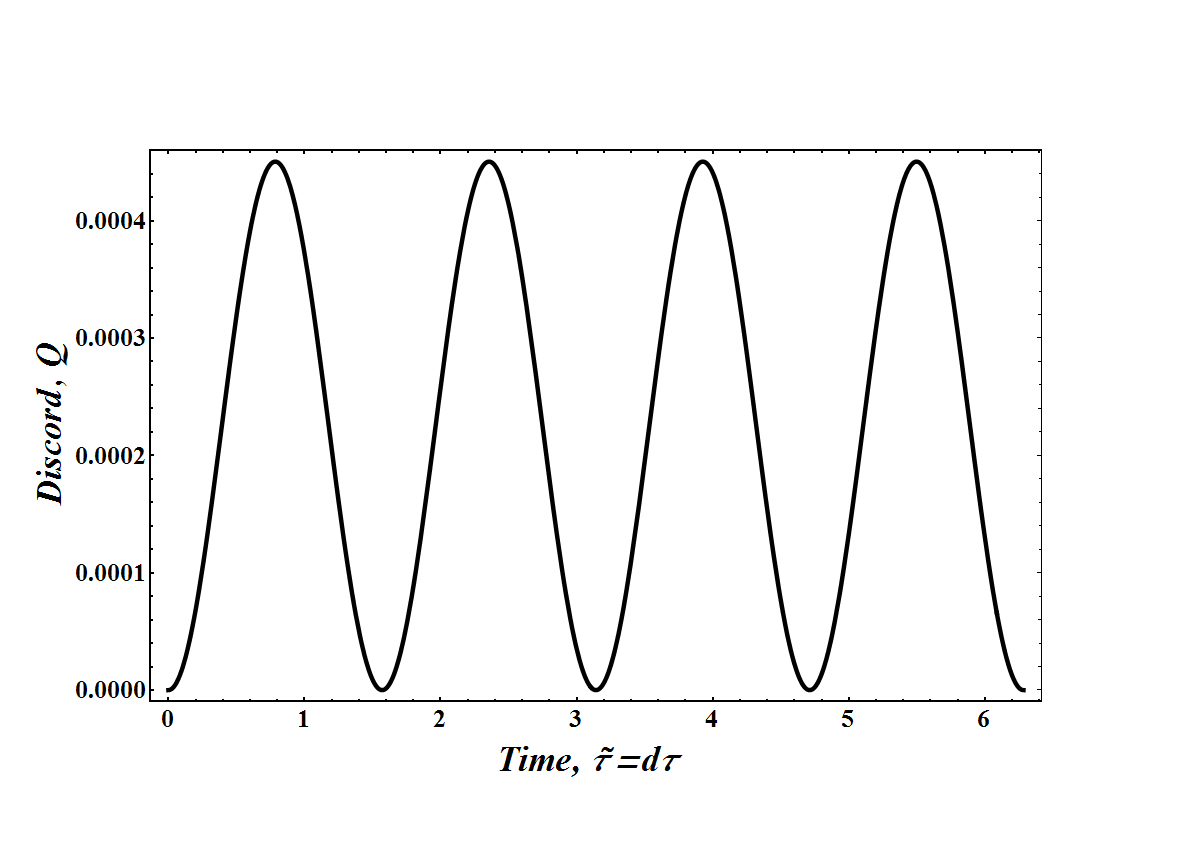}
	\end{tabular}
	\end{center}
   \caption[example] 
   { \label{fig:risunok2} 
The dependence of the quantum discord $Q$ on the dimensionless time $\bar{\tau}=d\tau$ of the preparation period of the MQ NMR experiment. According to Eq. \eqref{usloviyamin} $n_x=0,\; |n_y|=1,\; n_z=0;$ $\alpha=\beta=\frac{1}{\sqrt{2}}, b=0.1<1$.}
   \end{figure} 
Then the quantum discord can be found by the standard method \cite{doronin2015}. The dependence of the quantum discord on the dimensionless time of the preparation period of the MQ NMR experiment is given in Fig. 2.

\section{Conclusions}

We consider MQ NMR dynamics of dimers and show that the MQ NMR coherences of the zeroth and second orders only emerge in dimer systems. We also explore quantum correlations for dimers. We have found a connection between entanglement and the intensity of the MQ NMR coherence of the second order. We have also investigate the quantum discord.

\acknowledgments 
 
The work is supported by the Russian Foundation of Basic Research (Grant 16-03-00056) and the Program of RAS ''Element base of quantum computers (Grant No. 0089-2015-0220).

We thank Professor E.B. Fel'dman for stimulating discussions.


\end{document}